\documentclass[smallextended]{svjour3}

\smartqed

\usepackage{amsmath}
\usepackage{amssymb}
\usepackage{graphicx}
\usepackage{dcolumn}
\usepackage{tabularx}
\usepackage{soul}
\usepackage{color}
\usepackage{bm}
\usepackage{latexsym}
\usepackage{subcaption}

\usepackage{multirow}
\usepackage{hyperref}
\usepackage[normalem]{ulem}

\begin{document}

\title{Local NMR relaxation of dendrimers in the presence of hydrodynamic interactions}
\date{\today}

\author{Maxim Dolgushev \and
Sebastian Schnell \and
Denis A. Markelov
}

\institute{M. Dolgushev \at
              Institute of Physics, University of Freiburg, Hermann-Herder-Str. 3, 79104 Freiburg, Germany \\
Institut Charles Sadron, Universit\'e de Strasbourg and CNRS, 23 rue du Loess, 67034 Strasbourg Cedex, France \\ 
              \email{dolgushev@physik.uni-freiburg.de}\\
              Tel.: +49-761-2037688\\
           \and
           S. Schnell \at
              Institute of Physics, University of Freiburg, Hermann-Herder-Str. 3, 79104 Freiburg, Germany
              \and
              D. A. Markelov \at
              St. Petersburg State University, 7/9 Universitetskaya nab., St. Petersburg, 199034, Russia \\
              St. Petersburg National Research University of Information Technologies, Mechanics and Optics (ITMO University), Kronverkskiy pr. 49, St. Petersburg, 197101, Russia
}

\date{Received: date / Accepted: date}

\maketitle

\begin{abstract}
We study the role of hydrodynamic interactions for the relaxation of segments' orientations in dendrimers. The dynamics is considered in the Zimm framework. It is shown that inclusion of correlations between segments' orientations plays a major role for the segments' mobility, that reveals itself in the NMR relaxation functions. The enhancement of the reorientation dynamics of segments due to the hydrodynamic interactions is more significant for the inner segments. This effect is clearly pronounced in the reduced spectral density $\omega J(\omega)$, whose maximum shifts to higher frequencies when the hydrodynamic interactions are taken into account.
\end{abstract}

\section{Introduction}

Dendrimers are treelike macromolecules with a regular branching. Because of their unique architecture, there are plenty of applications of these macromolecules \cite{bosman99,grayson01,lee05}. Exemplarily, dendrimers can be used as drug delivery systems \cite{gillies05,hsu16}, nanoscale catalysts \cite{astruc01,caminade16}, rheology modifiers \cite{wang02,hajizadeh14}, contrast agents \cite{wiener94,sun16}, to name only a few of possible applications. Clearly, for some of these applications the local dynamical behavior is of a great importance.

Recently, much of attention has been attracted to the local dynamics in dendrimers, both in theory (analytic theory \cite{kumar13b,markelov14,grimm16} and computer simulations \cite{markelov15,shavykin16,markelov16}) and experiments \cite{pinto13a,pinto13b,hofmann15,mohamed15}, see also recent review \cite{markelov17}. Especially the NMR relaxation experiments have remarkably advanced this field. In particular, it was found that the methodology used for the analysis of mobility based on the spin-lattice relaxation  time $T_1$ has to be reexamined for dendrimers \cite{pinto13a}. Unlike for linear polymer chains, for dendrimers the mobility cannot be assessed based on a single frequency measurement only. In case of dendrimers, which possess a very broad relaxation spectrum, the $T_1$-function reveals its non-monotonous behavior. Therefore  measurements of $T_1$ at different frequencies or investigations of the spin-spin relaxation time $T_2$ are necessary \cite{pinto13a}.

The description of this remarkable local dynamics of dendrimers has been provided by the theory \cite{markelov14}. Unlike for linear chains, the dendrimer possesses exponentially growing relaxation times related to the dynamics of its large subbranches. As has been recently shown in Ref. \cite{markelov14}, in order to see these times in the local characteristics, one has to include local correlations between segments (i.e. to consider the so-called \textit{semiflexible dendrimers}). Nevertheless, the theoretical study of Ref. \cite{markelov14} did not include hydrodynamic interactions (HI), although the experiments typically deal with dendrimers in a solvent \cite{pinto13a,pinto13b,chai01,sagidullin03,malveau03,markelov10,markelov16b}. It is important to mention the work of Ref. \cite{kumar13b}, which studied the NMR relaxation functions for semiflexible dendrimers in solution averaged of the whole dendrimer structure. However, in the present work we are interested in the dependence of the NMR functions on the segments' location, bearing in mind the experiments of Refs. \cite{pinto13a,pinto13b}. As we proceed to show, the NMR functions of semiflexible dendrimers in solution strongly depend on the segments' location, although HI typically enhance mobility.
 
The paper is structured as follows: Sec.~\ref{model} represents the theory of the local dynamics of semiflexible dendrimers in solution. In Sec.~\ref{results} we provide and discuss our results. The paper ends with conclusions (Sec.~\ref{conclusions}). 

\section{Theory}\label{model}

\subsection{The model}

A dendrimer is a polymer with a regular treelike structure. To construct a dendrimer we start with a central bead to which we attach $f$ beads. This creates a dendrimer of generation $G=1$. The procedure is continued by attaching $f-1$ new beads to the peripheral beads, which creates a dendrimer of generation $G=2$. Iterating the previous step will increase the generation of the dendrimer by one for each iteration. We focus here on dendrimers with $f=3$ and various generation $G=3,\dots ,5$. Also, we enumerate the segments (i.e. springs) belonging to the same shell by $g$, starting with the segments attached to the core. (Note that for dendrimers the segments belonging to the same shell $g$ are equivalent.) Moreover, as we will show below, for the analysis of the segments' dynamics it is practical to introduce also an enumeration of shells from the periphery, i.e. to use $m\equiv G-g$. In this notation $m=0$ will indicate the outer (peripheral) shell. 

\begin{figure}[!t]
\begin{center}
\includegraphics[width=8.5cm] {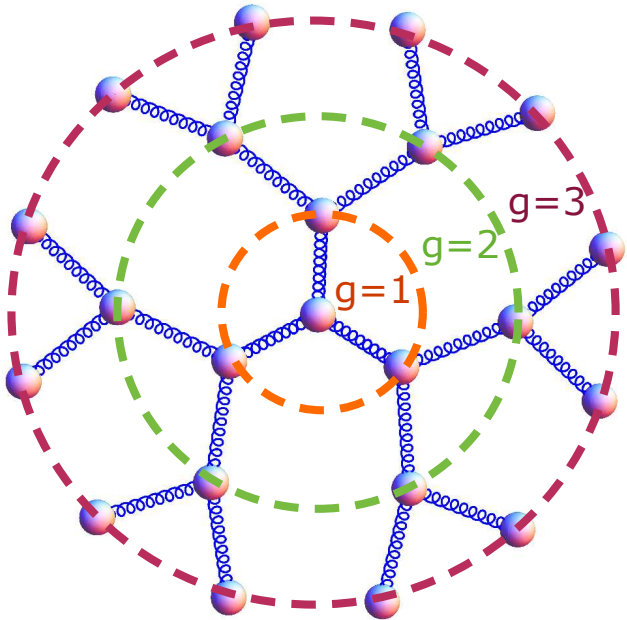}
\end{center}
\caption{Schematic representation of a dendrimer of generation $G=3$ and functionality $f=3$. The segments are represented by springs and the beads by spheres. The dashed circles indicate different shells numbered by $g$. Another enumeration scheme counts shells from the periphery, $m\equiv G-g$.}\label{fig1}
\end{figure}

In this work we consider dendrimers being constructed of identical beads, which are connected via  harmonic springs, called also segments. All springs have the same spring constant $K$ and the same mean-square length $l^2$. In this way, the structure of a dendrimer can be represented by a set of the beads' position vectors $\{\mathbf{r}_i\}$. For two connected beads (say, $i$ and $j$) we define an oriented segment $\mathbf{d}_a=\mathbf{r}_i-\mathbf{r}_j$.  To model semiflexibility we consider the orientations of segments to be correlated following the general framework of Ref. \cite{dolgushev09a} (that stems from earlier works \cite{bixon78,gotlib80,guenza92,winkler94,laferla97,vonferber02}). In the model of freely-rotating segments \cite{doi88} one has the following constraints. The mean-square segment lengths are fixed $\langle \mathbf{d}_a \cdot \mathbf{d}_a\rangle=l^2$. Two adjacent segments, say $a$ and $b$, fulfill  $\langle \mathbf{d}_a \cdot \mathbf{d}_b\rangle=\pm l^2q$, where the plus is for head-to-tail orientations and the minus otherwise. The semiflexibility parameter $q$ varies from $0$ to $1/(f-1)$, so for $f=3$ from $0$ to $1/2$, see Ref.\cite{mansfield80}. A flexible dendrimer has $q=0$, whereas a semiflexible dendrimer (SD) has $q>0$ (in this work we choose $q=0.45$ for SD). Finally, any two non-adjacent segments $a$ and $c$ are connected in a dendrimer through the unique path ($b_1,\cdots , b_k$). For them $\langle \mathbf{d}_a \cdot \mathbf{d}_c \rangle=\langle \mathbf{d}_a \cdot \mathbf{d}_{b_1} \rangle \langle \mathbf{d}_{b_1} \cdot \mathbf{d}_{b_2}\rangle \dots \langle \mathbf{d}_{b_k} \cdot \mathbf{d}_c \rangle l^{-2k}$ holds. For a dendrimer we can write these relations in a compact form, see Ref.~\cite{markelov14}
\begin{equation}\label{Korrelation da und db}
\langle \mathbf{d}_a\cdot \mathbf{d}_b\rangle=(-1)^{k-s}q^kl^2\,,
\end{equation}
where $k$ is the amount of beads along the unique path from $\mathbf{d}_a$ to $\mathbf{d}_b$ and $s$ equals the number of head-to-tail connections along this path. Here we choose the vectors representing the segments to point away from the core and therefore adjacent segments are oriented tail-to-tail for  segments belonging to the same shell and head-to-tail for the segments from different shells.

Equation~\eqref{Korrelation da und db} represents the covariance matrix of the multivariate Gaussian distribution for segments $\{\mathbf{d}_a\}$. Thus, given that all segments have a zero mean, the Gaussian distribution for $\{\mathbf{d}_a\}$ is fully determined through Eq.~\eqref{Korrelation da und db}. This distribution yields the Boltzmann distribution, $\exp(-V_{\text{SD}}(\{\mathbf{d}_a\})/k_BT)/Z$, with the potential energy
\begin{equation}\label{Potential}
V_{\text{SD}}(\{\mathbf{d}_a\})=\frac{K}{2}\sum_{a,b}^{}W_{ab}\mathbf{d}_a\cdot \mathbf{d}_b\,,
\end{equation}
where $K=3k_BT/l^2$ is the entropic spring constant and the matrix $\mathbf{W}=\{ W_{ab} \}$ is related to Eq.~\eqref{Korrelation da und db} by
\begin{equation}\label{matrix W}
\langle \mathbf{d}_a \cdot \mathbf{d}_b \rangle=l^2(\mathbf{W}^{-1})_{ab}\,.
\end{equation}
Hence in order to obtain the potential described by Eq.~\eqref{Potential} one has to invert the matrix of correlations represented by Eq.~\eqref{Korrelation da und db}. Strikingly, in the model considered here for any tree-like architecture the matrix $\mathbf{W}$ is known analytically, see Ref. \cite{dolgushev09a}. It turns out that the matrix $\mathbf{W}$ is very sparse \cite{dolgushev09a}. Elements involving nonadjacent segments vanish. For dendrimers the diagonal elements of $\mathbf{W}$ can take only two different values \cite{dolgushev09b},
\begin{align*}
&(-1+q)/(-1+q+2q^2)
\end{align*}
for peripheral segments and 
\begin{align*}
&(-1+q-2q^2)/(-1+q+2q^2)
\end{align*}
for non-peripheral segments. The off-diagonal elements involving adjacent segments are \cite{dolgushev09b}
\begin{align*}
\pm q/(-1+q+2q^2),
\end{align*}
where the plus sign stands for segments oriented as head-to-tail and the minus sign for other orientations.

The potential energy $V_{\text{SD}}(\{\mathbf{d}_a\})$ of Eq.~\eqref{Potential} can be transformed from segments into beads' position variables, $V_{\text{SD}}(\{\mathbf{r}_k\})$. This can be done using
\begin{equation}\label{Transformation}
\mathbf{d}_a\equiv \sum_{k}^{ }(\mathbf{G}^T)_{ak}\mathbf{r}_k\, .
\end{equation}
\newline
The matrix $\mathbf{G}$ is the so-called incidence matrix of the graph theory \cite{biggs93} and $^T$ denotes the transposition operation. The elements of $\mathbf{G}=(G_{ia})$, corresponding to the segment $a$, are either $G_{ja}=-1$ or $G_{ia}=1$ if the segment $a$ is orientated from bead $i$ to bead $j$ and zero otherwise. The transformation of Eq.~\eqref{Transformation} leads to
\begin{equation}\label{Potential_r}
V_{\text{SD}}(\{\mathbf{r}_k\})=\frac{K}{2}\sum_{i,j}^{}A_{ij}\mathbf{r}_i\cdot \mathbf{r}_j\,,
\end{equation}
with $\mathbf{A}=(A_{ij})$:
\begin{equation}\label{A=GWG}
\mathbf{A}=\mathbf{GWG}^T\, ,
\end{equation}
where $\mathbf{A}$ is the so-called dynamical matrix. Its elements are known analytically and listed elsewhere (see Ref. \cite{dolgushev09a} for general treelike structures and Ref. \cite{fuerstenberg12} for dendrimers).  Note that $\mathbf{A}$ and $\mathbf{W}$ are both square and symmetric but have different dimensions.  The dimensions of $\mathbf{A}$ are $N\times N$ and of $\mathbf{W}$ are $(N-1)\times (N-1)$, where $N$ stands for the number of beads. On the other hand, matrix $\mathbf{A}$ contains one zero eigenvalue $\lambda_1=0$ and $\mathbf{W}$ does not have vanishing eigenvalues, so that the rank of both matrices is equal to $(N-1)$.

For further calculations it is practical to use the normal modes $\{ \mathbf{u}_i\}=\{ u_{xi},u_{yi},u_{zi}  \}$ that are related to $\{\mathbf{r}_i\}=\{ r_{xi},r_{yi},r_{zi}  \}$ by 
\begin{equation}\label{Transform from r to u}
r_{\alpha i}(t)=\sum_{j}^{}Q_{ij}u_{\alpha j}(t)\, .
\end{equation}
Here $\mathbf{Q}=(Q_{ij})$ is constructed from orthonormal eigenvectors of $\mathbf{A}$, i.e., $\mathbf{Q}$ diagonalizes  $\mathbf{A}$:
\begin{equation}\label{A diagonal}
\mathbf{Q}^{-1}\mathbf{A\,Q}=\text{Diag}(\lambda_1,\,\dots,\lambda_N)\,.
\end{equation}
With this the potential energy of Eq.~\eqref{Potential_r} can be rewritten as
\begin{equation}\label{Potential_u}
V_{\text{SD}}(\{\mathbf{u}_n\})=\frac{K}{2}\sum_{i=2}^{N}\lambda_i\,\mathbf{u}_i\cdot \mathbf{u}_i\,,
\end{equation}
where we have used that $\lambda_1=0$.

\subsection{Hydrodynamic interactions}

In this work we study the dynamics of semiflexible dendrimers in a solvent, where the beads  experience HI. Each moving bead in the solvent creates a fluid current around itself and the surrounding beads are affected by this current. Following the Zimm-picture \cite{zimm56,doi88} HI are modeled by the Oseen tensor \cite{teraoka02},
\begin{equation}\label{H depends on r}
\widehat{\mathbf{H}}_{ij}=\mathbf{I}\delta_{ij}+\frac{3}{4}\frac{l\zeta_r}{R_{ij}}\left( \frac{\mathbf{R}_{ij}\otimes \mathbf{R}_{ij}}{R^2_{ij}} + \mathbf{I}\right)\left( 1-\delta_{ij} \right)\,,
\end{equation}\newline
where $R_{ij}=|\mathbf{R}_{ij}|=|\mathbf{r}_i-\mathbf{r}_j|$ and $\zeta_r=a/l$ is the coefficient related to the bead radius $a$. In this work we use the traditional value $\zeta_r=0.25$, as in Refs. \cite{osaki72a,osaki72b,guenza92,biswas01,kumar10,galiceanu14,galiceanu16}. This choice of $\zeta_r$ ensures the stability of dynamic quantities \cite{osaki72a,osaki72b,biswas01,galiceanu14}.

Based on the hydrodynamic tensor $\widehat{\mathbf{H}}$ and on the potential energy of Eq.~\eqref{Potential_r} we can construct a set of Langevin equations that describes the motion of beads. For, say, bead $i$ it reads 
\begin{equation}\label{Langevin with H}
\zeta\frac{\partial}{\partial t}\mathbf{r}_{ i}(t)=\sum\limits_{j=1}^{N}\widehat{\mathbf{H}}_{ij}\cdot\left(-K\sum\limits_{k=1}^{N}A_{jk}\mathbf{r}_{ k}+\mathbf{f}_{ j}(t)\right)\, 
\end{equation}
Here the left-hand side term represents friction force, i.e. $\zeta$ is the friction constant for a bead. The last term contains stochastic forces $\{\mathbf{f}_k(t)\}$, for which $\langle \mathbf{f}_k(t) \rangle=0$ and $\langle f_{\alpha k}(t)f_{\beta m}(t^\prime) \rangle=2 [(\widehat{\mathbf{H}}^{-1})_{nm}]_{\alpha \beta} k_BT\zeta 
\delta(t-t^\prime)$ holds. Now, since $\widehat{\mathbf{H}}_{ij}$ depends on $\mathbf{r}_j$, Eq.\eqref{Langevin with H} is not linear and very difficult to solve. To overcome this problem one uses in the Zimm picture the so-called preaveraging approximation \cite{zimm56,doi88}, in which $\widehat{\mathbf{H}}_{ij}$ is replaced by its equilibrium average value, which we call $\mathbf{H}_{ij}$ in the following. The interbead distances are Gaussian distributed and  the Cartesian components of $\{\mathbf{r}_{k}\}$ are uncorrelated. With this one gets \cite{doi88}
\begin{equation}\label{Definition H}
\mathbf{H}_{nm}=(\delta_{nm}+\zeta_r \langle l/R_{nm} \rangle(1-\delta_{nm}))\mathbf{I}\equiv H_{nm}\mathbf{I} \, ,
\end{equation}
where $\mathbf{I}$ is the three dimensional identity tensor. Moreover, 
the vector $\mathbf{R}_{nm}$ connecting beads $n$ and $m$ obeys a Gaussian distribution, so that for $\mathbf{R}_{nm}$ generally holds
\begin{equation}\label{Definition R}
\langle R_{nm}^{-1} \rangle=\left(\frac{6}{\pi \langle R^{2}_{nm} \rangle}\right)^{1/2} \, ,
\end{equation}
i.e., Eq.~\eqref{Definition R} is independent of the polymeric topology \cite{guenza92,laferla97,biswas01,kumar10,galiceanu14,galiceanu16}. Furthermore, we note that the stationary distances $\langle R^2_{nm} \rangle$ are independent of the HI. Therefore, we can evaluate $\langle R^2_{nm}\rangle$ based on the eigenvalues and eigenvectors of $\mathbf{A}$. The answer reads 
\begin{equation}\label{R^2 definition}
\langle R^2_{nm} \rangle=l^2\sum_{k=2}^{N}\frac{b^2_{k_{nm}}}{\lambda_k} \, .
\end{equation}
where we have defined
\begin{equation}\label{b=Q-Q}\newline
b_{k_{nm}}=Q_{kn}-Q_{km} \, .
\end{equation}
In case of flexible dendrimers ($q=0$) Eq.~\eqref{b=Q-Q} results in the topological matrix (i.e. the matrix of topological distances between the beads) \cite{teraoka02}, for semiflexible dendrimers the $\langle R^2_{nm} \rangle$ are evaluated numerically.

\subsection{Local dynamics of dendrimers in solution}

First quantity of our interest is the single segment time-autocorrelation function defined by
\begin{equation}\label{Ma1}
M^a_1(t)\equiv  \langle \mathbf{d}_a(t)\cdot \mathbf{d}_a(0)  \rangle / l^{2}\, .
\end{equation}
This function can be found by solving the set Langevin equations Eq.~\eqref{Langevin with H}. However, since the product $\mathbf{H}\mathbf{A}$ is not symmetric, there are different left and right sided eigenvectors. Therefore, we will work here in the symmetrized picture, see e.g. Ref.~\cite{chen99}, in which the matrix describing the set of Langevin equation is symmetric. Using the Cholesky decomposition of $\mathbf{H}=\mathbf{CC}^T$, the Langevin equation, Eq.~\eqref{Langevin with H}, under preaveraging, say, for the $y$-component reads 
\begin{equation}
 \overset{\cdot}{y}_i(t)=-\frac{K}{\zeta}\sum\limits_{j=1}^{N}(\mathbf{CC}^T\mathbf{A})_{ij}y_j(t)+\frac{1}{\zeta}\sum\limits_{j=1}^{N}(\mathbf{CC}^T)_{ij}f_j(t)\,.
 \end{equation}\newline
 Multiplication with $\mathbf{C}^{-1}$ from the left side leads to
 \begin{equation}\label{Langevin to solve}
 \sum\limits_{i=1}^{N}(\mathbf{C}^{-1})_{ki}\overset{\cdot}{y}_i(t)=-\frac{K}{\zeta}\sum\limits_{j=1}^{N}(\mathbf{C}^T\mathbf{A}\mathbf{CC}^{-1})_{kj}y_j(t)+\frac{1}{\zeta}f_k^\prime(t)\,,
 \end{equation}
 where we set $f_k^\prime(t)=\sum\limits_{j=1}^{N}(\mathbf{C}^T)_{kj}f_j(t)$, for which now $\langle \mathbf{f}^\prime_k(t) \rangle=0$ and $\langle f^\prime_{\alpha k}(t)f^\prime_{\beta m}(t^\prime) \rangle=2 k_BT\zeta \delta_{km}\delta_{\alpha \beta}\delta(t-t^\prime)$ hold. Moreover, from the symmetry of $\mathbf{A}$ follows the symmetry of $\mathbf{C}^T\mathbf{AC}$. Hence we can find an orthogonal matrix $\tilde{\mathbf{Q}}= \{\tilde{Q}_{km}\} $ such that \begin{equation}\label{Diagonal}
\tilde{\mathbf{Q}}^{-1}\mathbf{C}^T\mathbf{AC}\tilde{\mathbf{Q}}=\text{Diag}(\tilde{\lambda}_1,\dots,\tilde{\lambda}_N)\,,
\end{equation}\newline
where the $\{\tilde{\lambda}_i\}$ are the eigenvalues of $\mathbf{C}^T\mathbf{AC}$, including the eigenvalue $\tilde{\lambda}_1=0$ (these eigenvalues are the same as those of $\mathbf{H}\mathbf{A}$).
In this way the matrix $\tilde{\mathbf{Q}}$ leads to a transformation from $\{\mathbf{r}_k\}$ to $\{\mathbf{\tilde{u}}_m\}$. From Eq. \eqref{Diagonal} follows \newline
\begin{equation}\label{Cy}
\sum\limits_{i=1}^{N}(\mathbf{C}^{-1})_{ki}y_i(t)=\sum\limits_{m=1}^{N}\tilde{Q}_{km}\tilde{u}_m(t)\,.
\end{equation}
With this transformation we obtain orthogonal eigenmodes $\{\mathbf{\tilde{u}}_m\}$ whose correlation functions read
\begin{equation}\label{<u*u>}
\langle \tilde{u}_{\alpha k}(t)\tilde{u}_{\beta m}(0) \rangle=\frac{l^2 \delta_{\alpha \beta}\delta_{km}\exp(-\tilde{\lambda}_mt/\tau_0)}{3\tilde{\lambda}_m} \,.
\end{equation}

Now, multiplying of Eq.~\eqref{Cy} from the left side by $\mathbf{C}$ leads to
\begin{equation}
y_n(t)=\sum\limits_{i=1}^{N}(\mathbf{CC}^{-1})_{ni}y_i(t)=\sum\limits_{m=1}^{N}(\mathbf{C}\tilde{\mathbf{Q}})_{nm}\tilde{u}_m(t)\,.
\end{equation}\newline
With this and from Eq.\eqref{Transformation} we get for the $y$-component of the segment $\mathbf{d}_a$
\begin{equation}\label{transform d-->u HI}
d_{a,y}(t)=\sum\limits_{n=1}^{N}(\mathbf{G}^T)_{an}y_n(t)=\sum\limits_{m=1}^{N}(\mathbf{G}^T\mathbf{C}\tilde{\mathbf{Q}})_{am}\tilde{u}_m(t)\,,
\end{equation}\newline
from which based on Eq.~\eqref{<u*u>} the single segment time-autocorrelation function in presence of HI follows:
\begin{equation}\label{M1a hydro}
M^a_1(t)=\sum\limits_{j=2}^{N}[(\mathbf{G}^T\mathbf{C}\tilde{\mathbf{Q}})_{aj}]^2\frac{\text{exp}[-\tilde{\lambda}_jt/\tau_0]}{\tilde{\lambda}_j}\,.
\end{equation}

Now, the function $M^a_1(t)$ is connected with the second Legendre polynomial
\begin{equation}\label{Pa2}
P^a_2(t)\equiv \frac{1}{2} \left(3\left\langle \frac{(\mathbf{d}_a(t)\cdot \mathbf{d}_a(0))^2}{|\mathbf{d}_a(t)|^2| \mathbf{d}_a(0))|^2}\right\rangle-1\right).
\end{equation}
For Gaussian-distributed $\{\mathbf{d}_a\}$, $P^a_2(t)$ can be expressed analytically from $M^a_1(t)$ \cite{khazanovich63,perico85}. The result reads \cite{perico85}:
\begin{equation}\label{P2Guenza}
   P_2^{a}(t)=1-3\left\{x^2-\frac{\pi}{2}x^3\left[1-\frac{2}{\pi}\arctan(x)\right]\right\},
 \end{equation}
where $x=\sqrt{1-(M_1^a(t))^2}/M_1^a(t)$. The Fourier transform of the second Legendre polynomial $P^a_2(t)$, the so-called spectral density
\begin{equation}\label{J}
J(\omega)=\int P^a_2(t)\,e^{-i\omega t} \mathrm{d}t,
\end{equation} 
is the fundamental quantity for determination of the NMR relaxation functions, such as $T_1$, $T_2$, and NOE, see e.g. Refs.~\cite{abragam61,kimmich04,kimmich12,chizhik14}.

\section{Results and Discussion}\label{results}

\subsection{Relaxation of segments}

We start our discussion with the results for function $M_1(t)$ calculated for dendrimers' segments. Here we follow the terminology introduced in Refs. \cite{gotlib07,markelov09a,markelov09b,markelov14}.

\begin{figure*}[!ht]
\begin{center}
\includegraphics[width=8.5cm] {fig2a.eps}
\includegraphics[width=9.2cm] {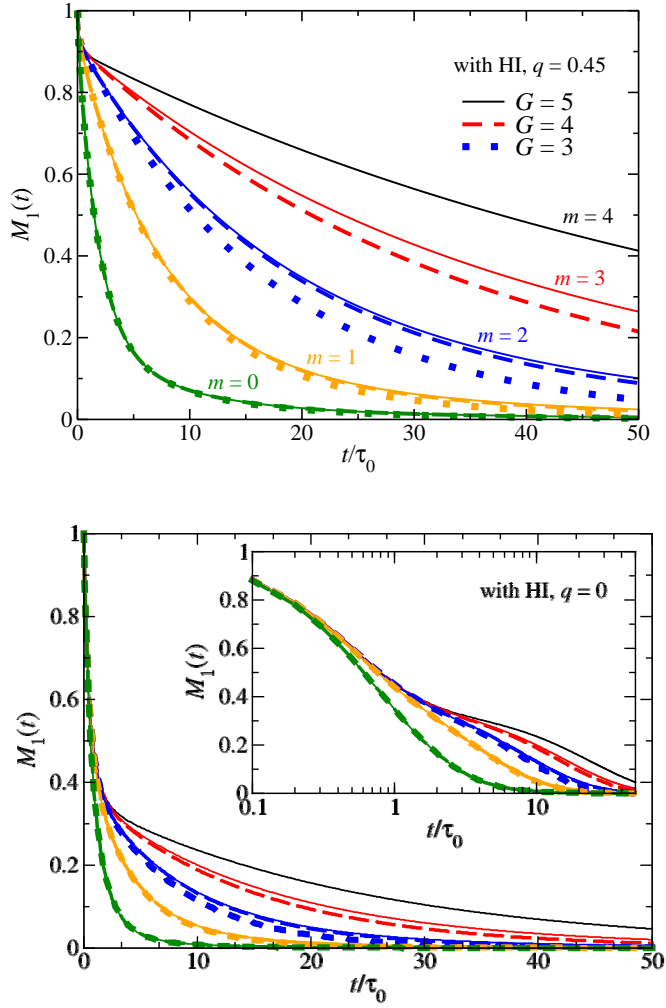}
\caption{(top) Temporal autocorrelation function $M_1^a(t)$ for segments of semiflexible dendrimers (of generation $G$) belonging to different shells counted from the periphery by $m$. (bottom) The same as for top figure, but for segments of  flexible dendrimers.}\label{fig2}
\end{center}
\end{figure*}

In Fig.~\ref{fig2} we display $M_1(t)$ for the segments belonging to different shells $m$ (the shells are counted from the periphery, so that index $m=0$ is related to the peripheral shell).  As can be inferred from the figure, for dendrimers the function $M_1(t)$ depends on $m$, but not on the dendrimer's generation $G$. The only considerable difference can be observed for $m=G-1$. These findings can be traced back to two major processes: (i) short-scale internal relaxation and (ii) relaxation of the branch originating from the labeled segment as a whole. The first process corresponds to the contribution of the internal relaxation modes. The ensuing part of the spectrum is located in a narrow region of the whole spectrum; it has a very weak dependence both on the size of the dendrimer $G$ and on the segment location $m$ (see also the inset to the bottom plot of Fig.~\ref{fig2}). Therefore this process can be described through an averaged relaxation time $\tau^{\mathrm{in}}$ related to the part of spectrum corresponding to the internal modes. This region of the spectrum is practically independent of $G$ and $m$.  On the contrary, the second process depends on the branch size, i.e. on index $m$ of its originating segment. Therefore the characteristic time $\tau_m^{\mathrm{br}}$ of this process grows with $m$.

Thus, the decay of the function $M_1(t)$ can be split on two regions: (i) the region of short times, where the function $M_1(t)$ has the same behavior for all $m$ characterized by the time $\tau^{\mathrm{in}}$ and (ii) the region of long times that one can characterize $M_1(t)$ by the time $\tau_m^{\mathrm{br}}$ that depends on $m$. As can be observed in Fig.~\ref{fig2}, for flexible dendrimers the initial region (i) dominates the dynamics. Inclusion of local stiffness leads to tremendous changes in the behavior of $M_1(t)$, so that the slow modes dominate the relaxation. This fact reflects  suppression of the local scale motions due to the introduced local bending stiffness.

As it was briefly mentioned above, the core segments (i.e. those with $m=G-1$) possess an exceptional behavior, especially for semiflexible dendrimers. One can observe that for the same value $m$ the functions with $m<G-1$ have a slower decay than those for $m=G-1$. This behavior can be traced back to the fact that for $m<G-1$ also the time $\tau_{m+1}^{\mathrm{br}}$ that is larger than $\tau_m^{\mathrm{br}}$, $\tau_{m+1}^{\mathrm{br}}>\tau_m^{\mathrm{br}}$, contributes. In case of $m=G-1$ the time $\tau_{m=G-1}^{\mathrm{br}}$ is the maximal relaxation time of the whole system, so that there are no larger times that can contribute.

\begin{figure*}[!ht]
\begin{center}
\includegraphics[width=8.5cm] {fig3a.eps}
\includegraphics[width=9.2cm] {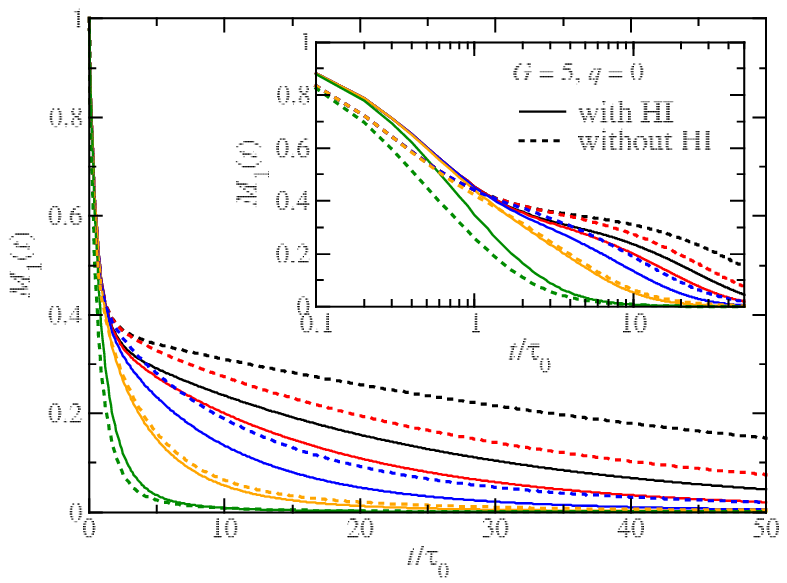}
\caption{Comparison of $M_1^a(t)$ calculated based on the models with HI (this work, Fig.~\ref{fig2}, $G=5$) and without HI (Ref.~\cite{markelov14}) for segments of semiflexible (top) and flexible (bottom) dendrimers.}\label{fig3}
\end{center}
\end{figure*}

We note that the behavior of $M_1(t)$ discussed above is in a qualitative agreement with previous theoretical works \cite{markelov16,markelov15,markelov14}. Hence we have shown that the inclusion of HI into the theoretical approach of Ref.~\cite{markelov14} does not change the qualitative behavior of the orientational mobility. These findings are also supported by the Brownian dynamics simulations \cite{lyulin04,markelov09a,markelov09b}. In order to look at the quantitative role of the HI, we compare our results with the theory that does not account for HI \cite{markelov14}, see Fig.~\ref{fig3}.

As can be observed in Fig.~\ref{fig3}, for large times, HI lead to a faster decay of $M_1(t)$, both for flexible and semiflexible dendrimers. For inner segments (indicated by larger $m$) this effect is even more pronounced than for more peripheral ones. This behavior corresponds to an effective decrease of the friction coefficient of beads. On the scale of large times the dynamics of a segment of $m$th shell involves relaxation of the whole subbranch that originates from this segment. For dendrimers of functionality $f=3$ this subbranch contains $2^{m+1}-1$ beads. Therefore the decrease of friction coefficient of a larger amount of beads leads to a stronger decrease for $\tau_m^{\mathrm{br}}$ corresponding to higher $m$. This consequently yields the quicker decay of $M_1(t)$ for higher $m$.

A final remark is related to short times. As can be observed for $m=0$, there is a very slight deviation from the general acceleration of the dynamics. This effect can be traced back to the fact that the small relaxation times and hence also $\tau^{\mathrm{in}}$ are related to the motion of neighboring beads in an antiphase manner \cite{cai97,gotlib02,fuerstenberg12}. HI rather decelerate such type of motions, thereby supporting the observed behavior on the short time scales. These findings are also supported by simulations \cite{lyulin04,markelov09a,markelov09b}, see e.g. Fig. 13 of Ref.~\cite{lyulin04}. 

\subsection{Spectral density}

\begin{figure*}[!t]
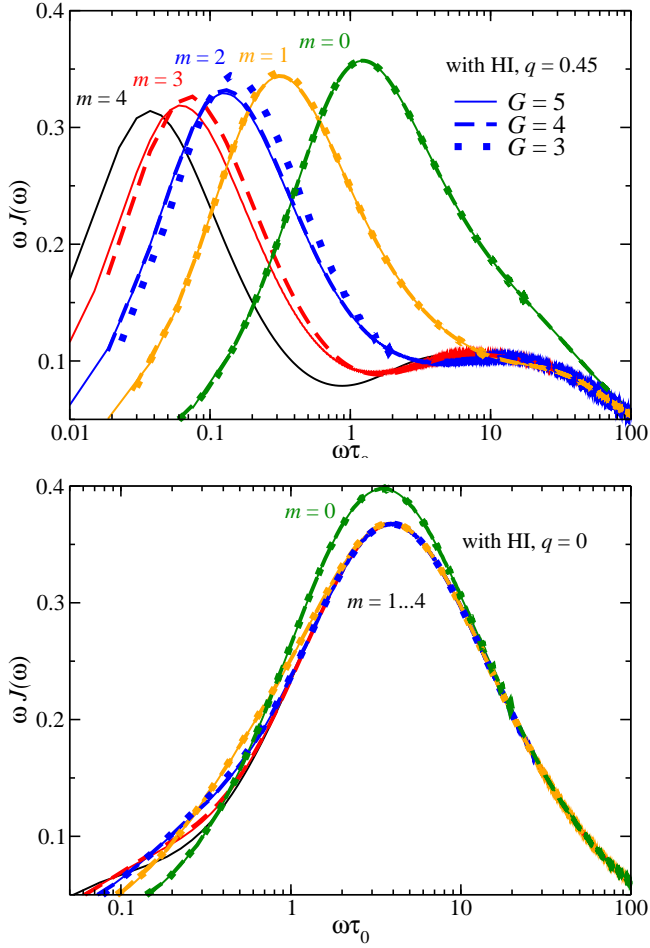

\begin{center}
\includegraphics[width=8.5cm] {fig4a.eps}
\includegraphics[width=8.5cm] {fig4b.eps}
\caption{(top) Reduced spectral density $\omega J(\omega)$for segments of semiflexible dendrimers (of generation $G$) belonging to different shells counted from the periphery by $m$. (bottom) The same as for top figure, but for segments of  flexible dendrimers.}\label{fig4}
\end{center}
\end{figure*}

\begin{figure*}[!t]
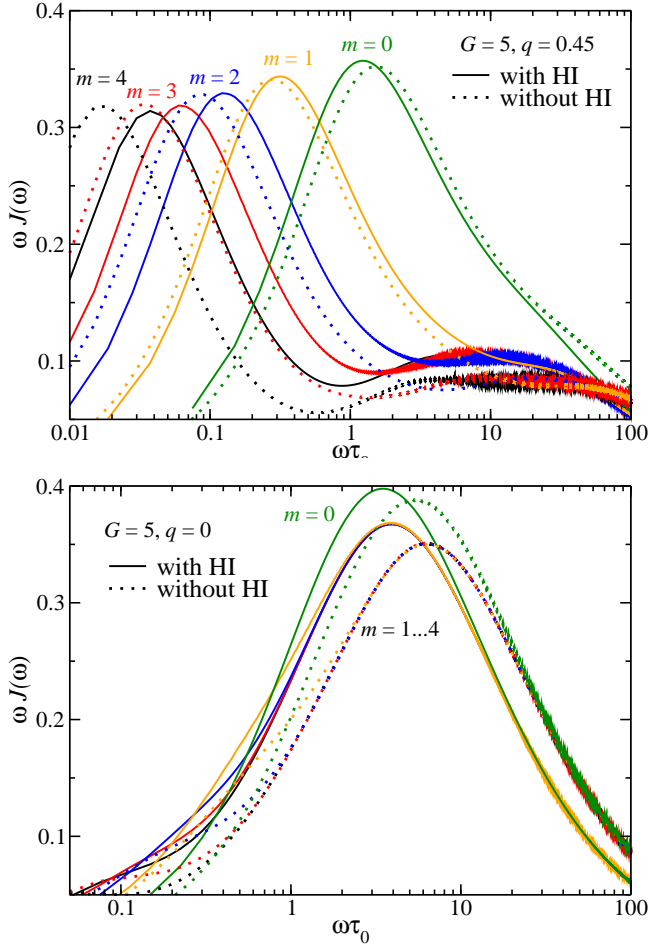

\begin{center}
\includegraphics[width=8.5cm] {fig5a.eps}
\includegraphics[width=8.5cm] {fig5b.eps}
\caption{Comparison of $\omega J(\omega)$ calculated based on the models with HI (this work, Fig.~\ref{fig4}, $G=5$) and without HI (Ref.~\cite{markelov14}) for segments of semiflexible (top) and flexible (bottom) dendrimers.}\label{fig5}
\end{center}
\end{figure*}

The reorientional autocorrelation function $M_1(t)$ discussed in the previous subsection is fundamental for calculation of spectral densities $J(\omega)$, see Eqs.~\eqref{P2Guenza} and \eqref{J}. The spectral densities of Figs.~\ref{fig4} and \ref{fig5} correspond to the functions  $M_1(t)$ of Figs.~\ref{fig2} and \ref{fig3}, respectively. 

In Fig.~\ref{fig4} we show the dependence of  spectral density on frequency for segments of flexible and semiflexible dendrimers in the presence of HI. As can be observed in Fig.~\ref{fig4}, for flexible dendrimers the spectral densities are practically independent of $m$ and $G$, with a little exception for $m=0$, for which the maximum of $\omega J(\omega)$ is slightly shifted towards low frequencies. This behavior stems from the fact that for $m=0$ mainly the local scale, inner modes contribute. Therefore the position of the maximum of $\omega J(\omega)$ is close to $1/\tau^{\mathrm{in}}$. In case of semiflexible dendrimer, the picture of $\omega J(\omega)$ displays striking deviations from $\omega J(\omega)$ of the flexible dendrimers. The maximum of $\omega J(\omega)$ is shifted towards low frequencies for segments that are closer to the core, i.e. for higher $m$. This shows that for semiflexible dendrimers the main contribution is related to the relaxation of the branch as a whole and the position of the maximum of $\omega J(\omega)$ is determined through the corresponding time $\tau_m^{\mathrm{br}}$ that is larger for higher $m$. We note that such differences in the behavior between flexible and semiflexible dendrimers were observed for the theoretical model that does not include HI \cite{markelov14}. Therefore in the following we discuss differences between the model of Ref. \cite{markelov14} and the present study by making a direct comparison between the ensuing functions $\omega J(\omega)$.

For a more detailed investigation of the influence of HI on the spectral density we compare it with that coming from the model without hydrodynamics \cite{markelov14}, see Fig.~\ref{fig5}. For semiflexible dendrimers, inclusion of HI leads to a shift of the maxima towards higher frequencies in comparison with the corresponding functions obtained in the model \cite{markelov14} that does not include HI. The reason for this tendency corresponds to decrease of $\tau_m^{\mathrm{br}}$ for the system with HI especially for higher $m$ (\textit{vide supra}). Interestingly, for $m=0$ there is a weak deviation from the general trend of $\omega J(\omega)$. Inclusion of HI leads to a slight increase of $\tau_0^{\mathrm{br}}$, therefore the corresponding function $\omega J(\omega)$ shifts towards lower frequencies. For flexible dendrimers this effect is even more pronounced, see Fig.~\eqref{fig5}(b). For this type of dendrimers one can observe this effect for all $m$, given that inclusion of HI leads to a slight growth of times of the inner spectrum (i.e. of $\tau^{\mathrm{in}}$) as well as of $\tau_0^{\mathrm{br}}$. Taking into account the results for the autocorrelation functions obtained in simulations \cite{markelov09a,markelov09b,lyulin04}, we can conclude that this effect is not an artifact of the viscoelastic model, but rather a feature of the dendritic structure that shows itself in its eigenmodes. We believe that further experimental and simulation studies can shed light on this specific feature of dendrimers. 

\section{Conclusions}\label{conclusions}

In this work we have studied the influence of hydrodynamic interactions on the reorientaional properties of segments in dendrimers. The hydrodynamic interactions have been modeled through the Ossen tensor, the dynamics of macromolecules has been considered in the Zimm picture. Two different viscoelastic models of dendrimers (flexible and semiflexible) have been examined. The results have been compared with those coming from the framework that does not include hydrodynamic interactions. The local relaxation of segments has been studied based on the temporal autocorrelation functions $M_1(t)$ and on the spectral density $J(\omega)$ that manifests the NMR relaxation experiments.

It has been shown that the inclusion of hydrodynamic interactions qualitatively conserves reorientational properties of segments, in particular, that functions $M_1(t)$ and $J(\omega)$ are determined by the remoteness of segments from the periphery. The presence of hydrodynamics leads to an acceleration of the decay of $M_1(t)$ and to a shift of the maximum of $\omega J(\omega)$ towards higher frequencies. This effect strengthens for more inner segments. An interesting exception from this behavior is provided by the peripheral segments. The obtained results are qualitatively supported by the computer simulations.

\begin{acknowledgements}

M.D. acknowledges the support through Grant No. GRK 1642/1 of the Deutsche Forschungsgemeinschaft. D.A.M. acknowledges the Russian Foundation for Basic Research (grant No. 14-03-00926) and the Government of the Russian Federation (grant 074-U01). 

\end{acknowledgements}

\bibliographystyle{spphys}

\begin{thebibliography}{10}
\providecommand{\url}[1]{{#1}}
\providecommand{\urlprefix}{URL }
\expandafter\ifx\csname urlstyle\endcsname\relax
  \providecommand{\doi}[1]{DOI \discretionary{}{}{}#1}\else
  \providecommand{\doi}{DOI \discretionary{}{}{}\begingroup
  \urlstyle{rm}\Url}\fi

\bibitem{bosman99}
A.W. Bosman, H.M. Janssen, E.W. Meijer, Chem. Rev. \textbf{99}(7), 1665 (1999)

\bibitem{grayson01}
S.M. Grayson, J.M.J. Frechet, Chem. Rev. \textbf{101}(12), 3819 (2001)

\bibitem{lee05}
C.C. Lee, J.A. MacKay, J.M.J. Fr{\'e}chet, F.C. Szoka, Nature Biotechnol.
  \textbf{23}(12), 1517 (2005)

\bibitem{gillies05}
E.R. Gillies, J.M.J. Frechet, Drug Discov. Today \textbf{10}(1), 35 (2005)

\bibitem{hsu16}
H.J. Hsu, J.~Bugno, S.r. Lee, S.~Hong, Wiley Interdisciplinary Reviews:
  Nanomedicine and Nanobiotechnology  (2016)

\bibitem{astruc01}
D.~Astruc, F.~Chardac, Chem. Rev. \textbf{101}(9), 2991 (2001)

\bibitem{caminade16}
A.M. Caminade, Chem. Soc. Rev. \textbf{45}, 5174 (2016)

\bibitem{wang02}
H.~Wang, G.P. Simon, C.~Hawker, C.~Tiu, Mater. Research Innov. \textbf{6}(4),
  160 (2002)

\bibitem{hajizadeh14}
E.~Hajizadeh, B.D. Todd, P.J. Daivis, J. Chem. Phys. \textbf{141}(19), 194905
  (2014)

\bibitem{wiener94}
E.~Wiener, M.~Brechbiel, H.~Brothers, R.L. Magin, O.~Gansow, D.~Tomalia,
  P.~Lauterbur, Magn. Res. Med. \textbf{31}(1), 1 (1994)

\bibitem{sun16}
W.~Sun, J.~Li, M.~Shen, X.~Shi, \emph{Dendrimer-Based Nanodevices as Contrast
  Agents for MR Imaging Applications} (Springer, 2016)

\bibitem{kumar13b}
A.~Kumar, P.~Biswas, Phys. Chem. Chem. Phys. \textbf{15}(46), 20294 (2013)

\bibitem{markelov14}
D.A. Markelov, M.~Dolgushev, Y.Y. Gotlib, A.~Blumen, J. Chem. Phys.
  \textbf{140}, 244904 (2014)

\bibitem{grimm16}
J.~Grimm, M.~Dolgushev, Phys. Chem. Chem. Phys. \textbf{18}(28), 19050 (2016)

\bibitem{markelov15}
D.A. Markelov, S.G. Falkovich, I.M. Neelov, M.Y. Ilyash, V.V. Matveev,
  E.~L{\"a}hderanta, P.~Ingman, A.A. Darinskii, Phys. Chem. Chem. Phys.
  \textbf{17}(5), 3214 (2015)

\bibitem{shavykin16}
O.V. Shavykin, I.M. Neelov, A.A. Darinskii, Phys. Chem. Chem. Phys.
  \textbf{18}(35), 24307 (2016)

\bibitem{markelov16}
D.A. Markelov, A.N. Shishkin, V.V. Matveev, A.V. Penkova, E.~L{\"a}hderanta,
  V.I.~Chizhik, Macromolecules \textbf{49}(23), 9247 (2016)

\bibitem{pinto13a}
L.F. Pinto, J.~Correa, M.~Martin-Pastor, R.~Riguera, E.~Fernandez-Megia, J. Am.
  Chem. Soc. \textbf{135}(5), 1972 (2013)

\bibitem{pinto13b}
L.F. Pinto, R.~Riguera, E.~Fernandez-Megia, J. Am. Chem. Soc. \textbf{135}(31),
  11513 (2013)

\bibitem{hofmann15}
M.~Hofmann, C.~Gainaru, B.~Cetinkaya, R.~Valiullin, N.~Fatkullin, E.A.
  R\"ossler, Macromolecules \textbf{48}(20), 7521 (2015)

\bibitem{mohamed15}
F.~Mohamed, M.~Hofmann, B.~P\"otzschner, N.~Fatkullin, E.A. R\"ossler,
  Macromolecules \textbf{48}(10), 3294 (2015)
  
\bibitem{markelov17}
D.A. Markelov, M.~Dolgushev, E.~L{\"a}hderanta, Annu. Rep. NMR Spectrosc.
  \textbf{91}, 1 (2017)

\bibitem{chai01}  
M. Chai , Y. Niu , W.J. Youngs , P.L. Rinaldi, J. Am. Chem. Soc. \textbf{123}(20), 4670 (2001)
  
\bibitem{sagidullin03}  
A. Sagidullin, V.D. Skirda, E.A. Tatarinova, A.M. Muzafarov, M.A. Krykin, A.N. Ozerin, B. Fritzinger, U. Scheler, Appl. Magn. Reson. \textbf{25}, 129 (2003)

\bibitem{malveau03}
C. Malveau, W.E. Baille, X.X. Zhu, W.T. Ford, J. Polym. Sci. Part B: Polym. Phys. \textbf{41}, 2969 (2003)

\bibitem{markelov10}
D.A. Markelov, V. V. Matveev, P. Ingman, M.N. Nikolaeva, E. L\"{a}hderanta, V.A.~Shevelev, N.I. Boiko. J. Phys. Chem. B \textbf{114}(12), 4159 (2010)

\bibitem{markelov16b}
D.A. Markelov, V. V. Matveev, P. Ingman, M.N. Nikolaeva, A.V. Penkova, E.~L\"{a}hderanta, N.I. Boiko, V.I. Chizhik. Sci. Rep. \textbf{6}, 24270 (2016)

\bibitem{dolgushev09a}
M.~Dolgushev, A.~Blumen, J. Chem. Phys. \textbf{131}, 044905 (2009)

\bibitem{bixon78}
M.~Bixon, R.~Zwanzig, J. Chem. Phys. \textbf{68}(4), 1896 (1978)

\bibitem{gotlib80}
Y.Y. Gotlib, Y.Y. Svetlov, Polym. Sci. U.S.S.R. \textbf{21}, 1682 (1980)

\bibitem{guenza92}
M.~Guenza, A.~Perico, Macromolecules \textbf{25}(22), 5942 (1992)

\bibitem{winkler94}
R.G. Winkler, P.~Reineker, L.~Harnau, J. Chem. Phys. \textbf{101}(9), 8119
  (1994)

\bibitem{laferla97}
R.~La~Ferla, J. Chem. Phys. \textbf{106}(2), 688 (1997)

\bibitem{vonferber02}
C.~von Ferber, A.~Blumen, J. Chem. Phys. \textbf{116}(19), 8616 (2002)

\bibitem{doi88}
M.~Doi, S.F. Edwards, \emph{The Theory of Polymer Dynamics} (Clarendon Press,
  1988)

\bibitem{mansfield80}
M.L. Mansfield, W.H. Stockmayer, Macromolecules \textbf{13}(6), 1713 (1980)

\bibitem{dolgushev09b}
M.~Dolgushev, A.~Blumen, Macromolecules \textbf{42}, 5378 (2009)

\bibitem{biggs93}
N.~Biggs, \emph{Algebraic graph theory} (Cambridge university press, 1993)

\bibitem{fuerstenberg12}
F.~F\"urstenberg, M.~Dolgushev, A.~Blumen, J. Chem. Phys. \textbf{136}, 154904
  (2012)

\bibitem{zimm56}
B.H. Zimm, J. Chem. Phys. \textbf{24}(2), 269 (1956)

\bibitem{teraoka02}
I.~Teraoka, \emph{Polymer Solutions} (Wiley Online Library, 2002)

\bibitem{osaki72a}
K.~Osaki, Macromolecules \textbf{5}(2), 141 (1972)

\bibitem{osaki72b}
K.~Osaki, J.L. Schrag, J.D. Ferry, Macromolecules \textbf{5}(2), 144 (1972)

\bibitem{biswas01}
P.~Biswas, R.~Kant, A.~Blumen, J. Chem. Phys. \textbf{114}(5), 2430 (2001)

\bibitem{kumar10}
A.~Kumar, P.~Biswas, Macromolecules \textbf{43}(17), 7378 (2010)

\bibitem{galiceanu14}
M.~Galiceanu, J. Chem. Phys. \textbf{140}(3), 034901 (2014)

\bibitem{galiceanu16}
M.~Galiceanu, A.~Jurjiu, J. Chem. Phys. \textbf{145}(10), 104901 (2016)

\bibitem{chen99}
Z.Y. Chen, C.~Cai, Macromolecules \textbf{32}, 5423 (1999)

\bibitem{khazanovich63}
T.~Khazanovich, Polym. Sci. U.S.S.R. \textbf{4}(4), 727 (1963)

\bibitem{perico85}
A.~Perico, M.~Guenza, J. Chem. Phys. \textbf{83}, 3103 (1985)

\bibitem{abragam61}
A.~Abragam, \emph{The principles of nuclear magnetism} (Oxford university
  press, 1961)

\bibitem{kimmich04}
R.~Kimmich, N.~Fatkullin, \emph{Polymer chain dynamics and NMR} (Springer,
  2004)

\bibitem{kimmich12}
R.~Kimmich, \emph{NMR: tomography, diffusometry, relaxometry} (Springer Science
  \& Business Media, 2012)

\bibitem{chizhik14}
V.I. Chizhik, Y.S. Chernyshev, A.V. Donets, V.V. Frolov, A.V. Komolkin, M.G.
  Shelyapina, \emph{Magnetic resonance and its applications} (Springer, 2014)

\bibitem{gotlib07}
Y.Y. Gotlib, D.A. Markelov, Polym. Sci. Ser. A \textbf{49}(10), 1137 (2007)

\bibitem{markelov09a}
D.A. Markelov, S.V. Lyulin, Y.Y. Gotlib, A.V. Lyulin, V.V. Matveev,
  E.~Lahderanta, A.A. Darinskii, J. Chem. Phys. \textbf{130}(4), 044907 (2009)

\bibitem{markelov09b}
D.A. Markelov, Y.Y. Gotlib, A.A. Darinskii, A.V. Lyulin, S.V. Lyulin, Polym.
  Sci. Ser. A \textbf{51}(3), 331 (2009)

\bibitem{lyulin04}
S.V. Lyulin, A.A. Darinskii, A.V. Lyulin, M.~Michels, Macromolecules
  \textbf{37}(12), 4676 (2004)

\bibitem{cai97}
C.~Cai, Z.Y. Chen, Macromolecules \textbf{30}, 5104 (1997)

\bibitem{gotlib02}
Y.Y. Gotlib, D.A. Markelov, Polym. Sci. Ser. A \textbf{44}(12), 1341 (2002)

\end{thebibliography}

\end{document}